\documentclass{nature}
\usepackage[utf8]{inputenc}
\usepackage{graphicx}
\usepackage{amsmath,amssymb,caption}
\usepackage{soul}
\usepackage[dvipsnames]{xcolor}
\usepackage[colorlinks=true]{hyperref}
\usepackage[utf8]{inputenc}
\usepackage[T1]{fontenc}
\usepackage{csquotes}
\usepackage{booktabs}
\usepackage{float} 
\newcommand{\ie}{\emph{i.e.}}

\newcommand{\eref}[1]{Eq.~(\ref{#1})}

\title{Measuring the impact of hits}
\author{Matúš Medo, Liudmila Rozanova}

\makeatletter
\let\saved@includegraphics\includegraphics
\AtBeginDocument{\let\includegraphics\saved@includegraphics}
\renewenvironment*{figure}{\@float{figure}}{\end@float}
\makeatother
\begin{document}
\maketitle

\begin{abstract}
Many real systems can be represented as growing networks where new nodes and links gradually emerge. The Barab\'asi-Albert model for growing networks, and many models inspired by it, are based on the idea that nodes compete for links. However, the strength and the very presence of this competition have not been tested. We propose a robust statistical approach to quantify how strongly nodes compete for links, and apply it to data from various real systems---commenting on online news and cinema attendance data. We find a range of possible behaviors, from the perfectly elastic case, where new entrants shape network growth in a way that leaves the rest of the system unaffected, to an intermediate case where new entrants measurably affect the rest. Perfect competition is never observed. These findings have direct implications for complex systems modeling and e-commerce applications.
\end{abstract}

\section*{Introduction}
Popularity dynamics and human attention are crucial drivers in many parts of our society~\cite{lorenz2019accelerating,pedersen2021political,caldarelli2026physics}. Of the many songs, movies, and scholarly papers created, only a few become successful and widely known, together with their authors. Because time and resources are limited, competition is an inherent part of popularity dynamics: songs compete for listeners' attention on online platforms, movies compete for viewers' attention in cinemas, and scholarly papers compete for researchers' attention to get cited and eventually help propel their authors' careers. In a world where collective attention is inherently limited~\cite{simon1996designing,weng2012competition}, successful items can impact the dynamics of the rest by reducing the collective attention that is left for them~\cite{wu2007novelty,gleeson2014competition,gleeson2016effects,lorenz2019accelerating}.

From the theoretical point of view, popularity dynamics has been traditionally modeled in two complementary ways. The Barab\'asi-Albert model~\cite{barabasi1999emergence}, one of the pivotal models in network science~\cite{newman2018networks}, together with its multiple generalizations, assumes that networks grow by adding nodes, and that each new node establishes a limited number of links to existing nodes in the network. Network nodes then effectively compete for the new links; success in this competition is shaped by preferential attachment, aging, and node fitness~\cite{medo2011temporal,golosovsky2018mechanisms}, among others~\cite{backstrom2006group,boguna2021network}. Assuming that nodes compete for incoming links has important consequences for the system's behavior. In~\cite{gleeson2014competition}, for example, it drives the system to criticality, resulting in very heavy tails of the popularity distribution. Another class of models considers decoupled dynamics where each item's success evolves independently of the others. The underlying mathematical models for such popularity dynamics include the Bass model~\cite{bass1969new}, reinforced Poisson process~\cite{shen2014modeling}, Hawkes process~\cite{rizoiu2017hawkes}, and attention decay~\cite{ojer2026modeling}, often leading to bursty behavior~\cite{karsai2018bursty}.

However, it remains unclear where real systems fall between these two extremes. Do items indeed compete for users' attention? To answer this fundamental question, we propose a principled statistical measurement. This measurement is based on a natural experiment~\cite{leatherdale2019natural} where the decay of attention in existing items is compared between two scenarios. In the baseline scenario, no new items arrive that could compete for user attention, or the arriving items fail to attract substantial attention. In the hit scenario, an item that immediately attracts considerable attention arrives and can influence the attention other items receive. By directly comparing attention decay between these two scenarios, we measure the elasticity of collective attention and thus infer the level of competition in the studied system. We verify the proposed measurement on a novel network growth model in which the level of competition can be tuned by setting the activity elasticity parameter between no competition, where activity/attention accommodate supply, and complete competition, where activity/attention are fixed.

We finally apply this methodology to three datasets: Commenting activity on 3,087 sports articles from the British Broadcasting Corporation (BBC), comments on 56,000 posts from the \emph{worldnews} community on Reddit, and 36 years of weekly US box office data. We find no statistically significant competition in both commenting datasets, suggesting that users are ready to spend more time reading and commenting when many interesting news items appear. In the cinema box office data, we observe only weak competition, despite the substantial time and money investment required for each cinema visit. 

In summary, we introduce a framework to distinguish between systems where resources (\emph{e.g.}, attention or funding) are exogenous and systems where they can vary endogenously. We find weak or absent competition in three real systems, which has direct commercial implications. The absence of competition in real systems also strongly supports theoretical models in which attention is not limited, but an endogenous property instead.

\section*{Results}

\subsection{Measuring the effect of hits on the other items}
As our time and energy are inherently limited, our collective attention is necessarily bounded. This idea is typically reflected in popularity dynamics models by assuming that the existing items compete for finite attention~\cite{medo2011temporal,weng2012competition,gleeson2014competition,lorenz2019accelerating}. Under the limited collective attention hypothesis, it is natural to expect that the appearance of a popular article on a news website (covering a major event in a presidential race or a natural disaster in a nearby country, for example) leads to a decrease in the attention given to the concurrent news. An alternative hypothesis is that such hits ``live on their own''. In this case, hits would increase overall activity on the platform, but they would not significantly affect attention to other articles.

We quantify the effect of a new hit item on the system's dynamics as follows. We measure the attention items attract by the number of new interactions they receive in two consecutive intervals. For the datasets used here, interactions are represented by comments for news items and by ticket sales for movies. The length of the time intervals, $\Delta t$, depends on the timescale of the system's dynamics. We use $\Delta t=10\,\text{minutes}$ for the news datasets and $\Delta t = 1\,\text{week}$ for the box office data (the latter choice helps smooth out cinema attendance, which varies strongly during the week). The consecutive time intervals are defined by the time separating them; we choose multiple time points, $t_j$, to map the system's dynamics over the full time range spanned by the data. The measured numbers of interactions of item $i$ are denoted $\Delta c_i^B(t_j)$ (\emph{before} $t_0$) and $\Delta c_i^A(t_j)$ (\emph{after} $t_0$), respectively. Because $\Delta t$ is much longer for the box office data, we adopt milder conditions for the baseline scenario: We consider all weeks in which no new and eventually successful movies (80th percentile and above) enter cinemas (see Methods for further details). Note that a similar before-and-after measurement design has been used in the past to assess the impact of a natural disaster on social network dynamics~\cite{phan2015natural}, for example.

We choose the reference time points in two distinct ways. In the first way, the \emph{baseline scenario}, we first identify all time points $t$ when no new items appeared during the period $[t-\Delta t, t + \Delta t]$. From these time points, we randomly choose $M$ to represent the dynamics of interactions. By construction, the number of interactions in those periods is unaffected by new competitors. As we will see later, the number of interactions decreases between the before and after windows, and this decrease is due solely to aging, which dictates that interest in items decreases with time. We then consider the \emph{hit scenario}, where we first compute the ranking of all items by the number of interactions they received in time $\Delta t$ since their appearance and choose $M'$ of them with the highest number of interactions; these items are referred to as \emph{hits} for brevity. The reference time points for the hit scenario are the hits' appearance times. In both cases, the evolution of the number of interactions is characterized by the multiplicative factor $\gamma$: When the expected activity of item $i$ in a window is $\mu_i$, in the subsequent window it is $\gamma\,\mu_i$. The previously documented overall decrease in the number of new comments over time~\cite{medo2022simple} suggests that $\gamma$ is typically smaller than one -- for this reason, we refer to $\gamma$ as the \emph{slowing factor}.

The maximum likelihood estimate of the slowing factor is
\begin{equation}
\label{gamma_est}
\hat\gamma=\frac{\sum_{i,j} \Delta c_i^A(t_j)}{\sum_{i,j} \Delta c_i^B(t_j)},
\end{equation}
as shown in Methods. In the hit scenario, the summation in \eref{gamma_est} excludes the hit article whose effect we aim to measure. The corresponding timescale of the exponential decay of activity is
\begin{equation}
\tau = -\Delta t / \ln\hat\gamma
\end{equation}
assuming $\hat\gamma<1$.

\begin{figure*}
\centering
\includegraphics[width=\textwidth]{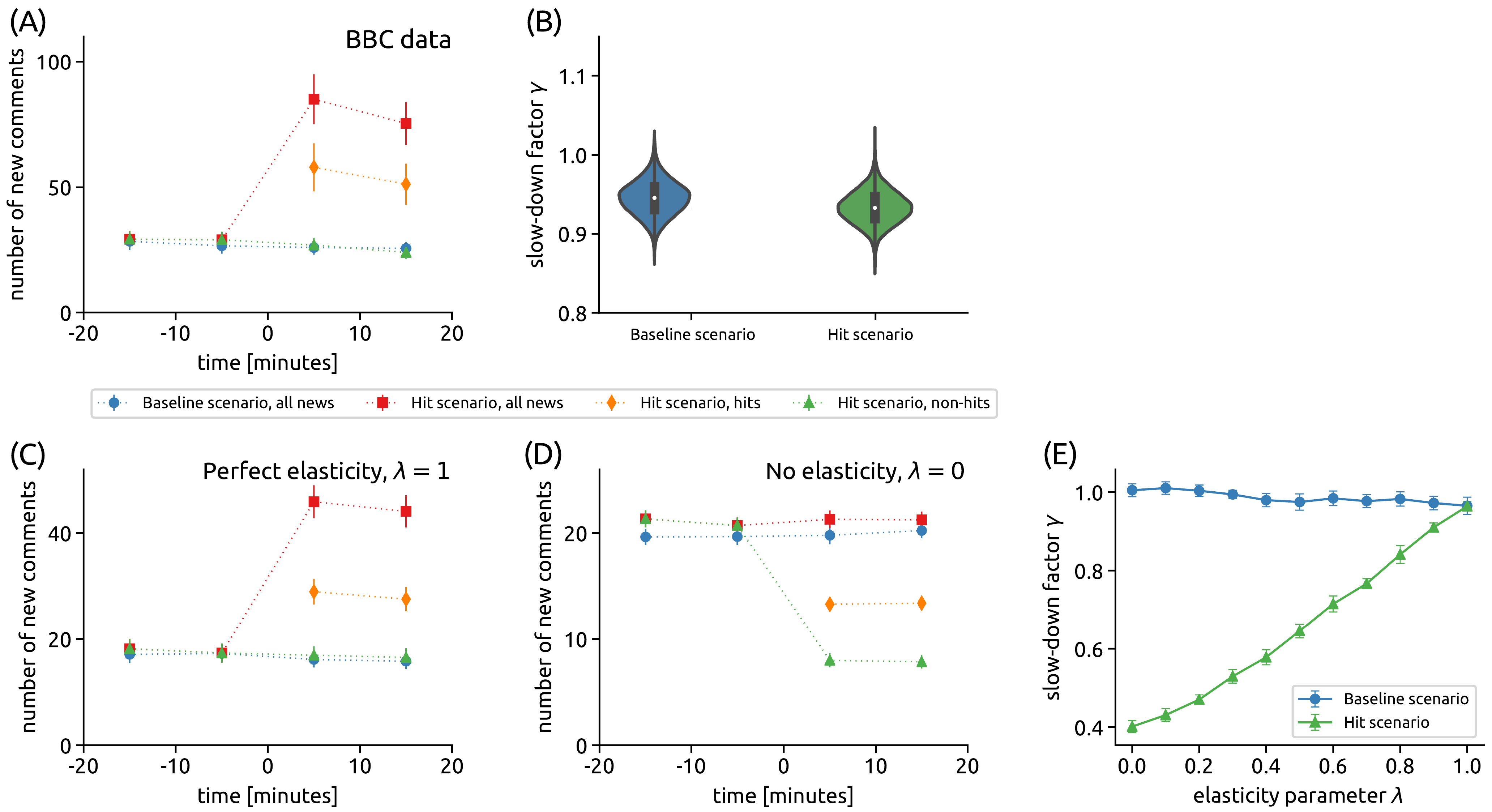}
\caption{\textbf{The effect of hits on the other articles in empirical data (top row) and in synthetic data with tunable elasticity of collective attention (bottom row).} In the baseline scenario, no new articles appear during the observed time period. In the hit scenario, a hit article (top 90th percentile) appears in the middle of the observed time period. (A) The average number of new comments in 10-minute time intervals in the baseline and hit scenarios. In the hit scenario, hit articles appear at time 0; error bars are twice the standard error of the mean. (B) The estimated slowing factors in the baseline and hit scenarios differ by only about 2\%, and the differences are not significant. (C, D) As panel (A) for the synthetic data with perfect elasticity of collective attention (C) and no elasticity (D), respectively. (E) Estimates of the slowing factor in the synthetic data in the baseline and hit scenarios for the full range of the elasticity parameter $\lambda$.}
\label{fig:fig1}
\end{figure*}

\subsection{BBC Sport comment sections}
In the baseline scenario, the number of interactions decreases with time in agreement with expectations~\ref{fig:fig1}A. The slowing factor is close to $0.95$ in the BBC data, corresponding to a characteristic decay timescale of $195$ minutes, consistent with results previously reported for this dataset in~\cite{medo2022simple}. The appearance of a hit dramatically increases the overall activity in the platforms (Fig.~\ref{fig:fig1}D). However, this increase is due to the hit item itself, as the average activity aimed at other existing items before and after the hit's appearance is similar, suggesting that the hit's influence on the rest of the system is negligible. In the BBC data, for example, hits receive, on average, twice as many comments as other items during the first $10$ minutes after their appearance. Despite the success of hits, we find that the average slowing factor in the hit scenario is only about 2\% smaller than in the baseline scenario for the BBC data (see Fig.~\ref{fig:fig1}B), and this difference is not statistically significant ($p=0.67$). This indicates that the emergence of a hit is associated with only a slightly faster decay of impact for the other articles.

Is the observed lack of effect of hits due to the fact that hits and ordinary articles attract substantially different audiences? To answer this question, one can divide both news articles and users into groups by the number of comments received and the number of articles commented on, respectively, and compare the number of links between various news article--user
groups with randomized data that preserve individual users' total activity and articles' total impact, as well as optionally also the time pattern of activity~\cite{ren2018randomizing}. A similar check of degree assortativity patterns for this dataset has been already done (see Figure 2 in~\cite{medo2022simple}) and showed little assortativity patterns. This indicates, among other things, that the hits do not reach high impact by activating a pool of little-active users but by nearly uniformly attracting attention from users of all activity levels.

\subsection{The effect of hits on the other articles is observable when user attention is not elastic}
A possible objection to the presented lack of effect of hits is that the proposed maximum-likelihood procedure might not be able to detect the effect of hits in cases where such an effect is veritably present. To rule out this possibility, we introduce a generalized dynamical model where the interplay between the user activity and the available articles is controlled by an elasticity parameter, $\lambda$ (see Materials and Methods) -- we refer to this model as the FAE model because it includes fitness, aging, and tunable elasticity. This model allows us to interpolate between two extreme worlds: A world where the collective attention can stretch without limitations in the presence of high-impact articles (perfect elasticity, $\lambda=1$, see Fig.~\ref{fig:fig1}C), and a world where the collective attention is fixed and independent of the presence of high-impact articles (no elasticity, $\lambda=0$, see Fig.~\ref{fig:fig1}D). In the perfect elasticity world, in line with our empirical findings, the appearance of a hit has no effect on the other articles (Fig.~\ref{fig:fig1}C), whereas in the no-elasticity world, the appearance of a hit is associated with a sharp reduction of the attention received by the other articles (Fig.~\ref{fig:fig1}D). Can our maximum-likelihood procedure detect the effect of hits in low-elasticity worlds?

To address this question, we calibrate the FAE model on the BBC data and create synthetic datasets across the full range of elasticity values (from no elasticity, $\lambda=0$, to perfect elasticity, $\lambda=1$; see Materials and Methods and Sec.~7 in the SI). As shown in Figure~\ref{fig:fig1}E, the estimated slow-down factor significantly differs between the baseline scenario and the hit scenario for most elasticity values, which shows that the proposed effect measurement can detect the effect of hits for all elasticity values except those close to $\lambda=1$. In particular, a non-significant difference ($p>0.05) $ between the slowing factors in observed in the baseline and hit scenarios emerges for $\lambda\gtrsim 0.96$. In light of this high sensitivity of the proposed slow-down factor to the elasticity of the system, we conclude that the empirical lack of effect of hits on the other articles suggests that the collective attention is highly elastic in real-world news outlets, deriving directly from the news articles present in the platform.

\begin{table}
\centering
\begin{tabular}{@{}lcc@{}}
\toprule
& \textbf{Hit scenario} & \textbf{Baseline scenario} \\
\midrule
Peak total (comments/min)        & 69.8  & 42.5  \\
Peak from hit post               & 57.7  & 18.6  \\
Peak background (non-hit posts)  & 12.0  & 23.9  \\
Baseline total                   & 20.2  & 13.5  \\
Total increase ratio (peak/base) & 3.45  & 3.15  \\
Background increase ratio        & 0.60  & 1.97  \\
\bottomrule
\end{tabular}
\caption{Aggregate activity metrics for hit vs.\ baseline
scenarios on Reddit.  The hit scenario shows a dramatic total
increase driven entirely by the hit post; the background rate
drops by ${\sim}40\%$ at the moment of peak.}
\label{tab:reddit:aggregate}
\end{table}

\begin{figure}
\centering
\includegraphics[width=\textwidth]{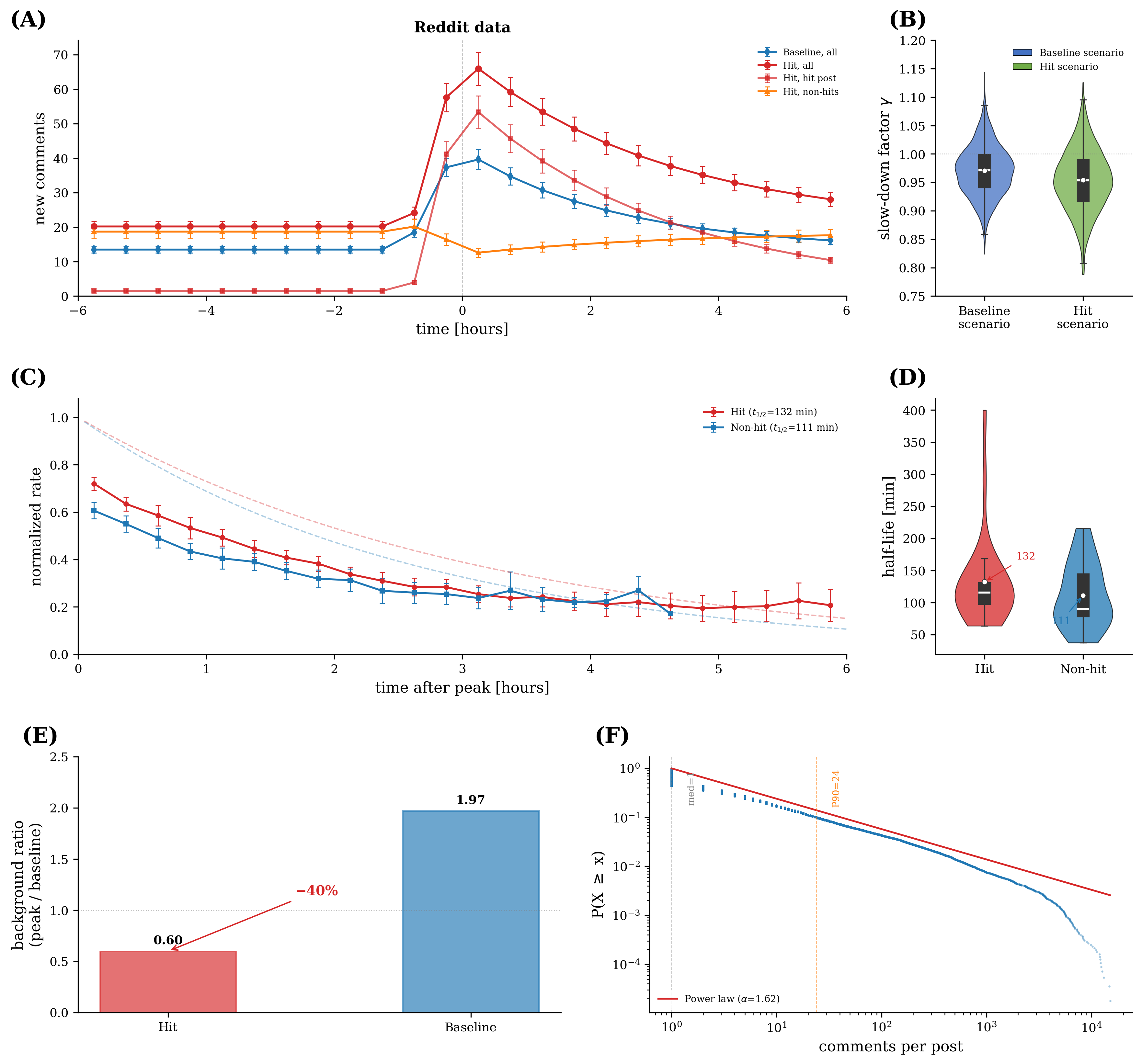}
\caption{\textbf{The effect of hits on the other posts in Reddit data.}
\textbf{(A)}~Average comment rate around hit appearance.  Total activity (red) surges due to the hit post (red squares); non-hit posts (orange triangles) remain near baseline (blue diamonds).  Error bars: $2\times$SEM.
\textbf{(B)}~Estimated slow-down factor $\hat\gamma$: difference ${\sim}2\%$, not significant.
\textbf{(C)}~Normalized comment-rate decay after peak.  Hit posts: $t_{1/2} = 132$~min; non-hit posts: $t_{1/2} = 111$~min.  Dashed lines: fitted exponentials.
\textbf{(D)}~Half-life distributions; means annotated.
\textbf{(E)}~Transient background suppression: background ratio drops to 0.60 at hit peak vs.\ 1.97 in baseline (${\sim}40\%$ drop).
\textbf{(F)}~Complementary cumulative distribution of comments per post; power law with $\alpha = 1.62$.}
\label{fig:reddit}
\end{figure}

\subsection{Reddit discussions}
We further analyze discussions of 56,000 posts on the social platform Reddit.com (see Methods). Reddit discussions are short-lived. On average, half of a post's lifetime comments arrive within the first 5~hours of its appearance, and 90\% of comments accumulate within about 14~hours. Direct modeling of the post-peak decline in activity revealed a consistent exponential decay pattern. While the total comment rate increases by a factor of ${\sim}3.5$ during hit appearances, this increase is driven almost entirely by the hit post itself (Table~\ref{tab:reddit:aggregate}).  The comment rate on non-hit posts remains near its pre-hit level of ${\sim}20$ comments per minute (Figure~\ref{fig:reddit}A, orange triangles). The estimated slow-down factors in the baseline ($\hat\gamma_B \approx 0.97$) and hit ($\hat\gamma_H \approx 0.95$) scenarios differ by only ${\sim}2\%$ (Figure~\ref{fig:reddit}B). This difference is not statistically significant, confirming no sustained competition effect.

\begin{figure*}
\centering
\includegraphics[scale=0.75]{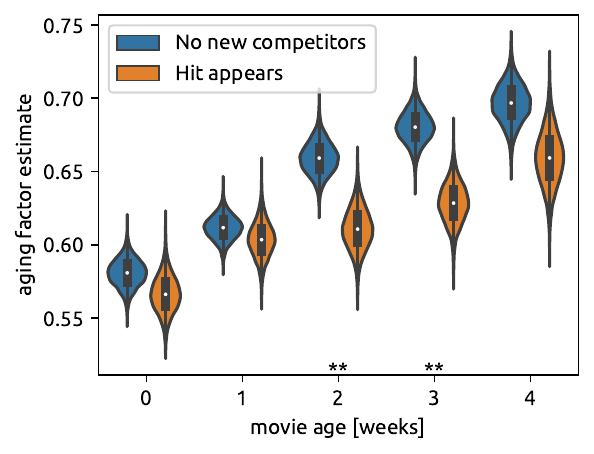}
\caption{\textbf{The effect of hits on the other movies in box office data.} We measured the slowing factor, $\gamma$, in movie revenues, $R$, between two consecutive weeks. Two scenarios are compared: (1) Baseline scenario where no new successful movies (total revenues above the 80th percentile) appear at $t+1$ (``No new competitors''), (2) Hit scenario where a hit movie (total revenues above the 99th percentile) appears at $t+1$ (``Hit appears'').}
\label{fig:moyo}
\end{figure*}

\subsection{Cinema box office data}
To complement two datasets from online platforms, we finally turn to the US box office data from years 1983--2018 (see Methods). We use here the movie gross weekly income as a proxy for the attention/number of interactions. While gross income is affected by increasing ticket prices in the long term, this does not pose complications to the evaluation of short-term impact of hits. For the cinema data, the slowing factor is much lower than before due to the substantially longer time window (1 week as opposed to minutes or tens of minutes before). The baseline scenario slowing factor is $0.58$, corresponding to a decay timescale of 12.9 days. Upon emergence of a hit, the aging factor decreases as expected, but the decrease is less than 0.02 and the difference is not statistically significant (Figure~\ref{fig:moyo}). Interestingly, the slowing factor differences between the baseline and hit scenario become significant later on, starting from week 3 after the movie has been released. Over the displayed 5-week period, during which movies typically reach the majority of their box office income, the compounded difference between the baseline scenario and the hit scenario is only about 5\%.

\begin{table}
\centering
\small
\begin{tabular}{@{}lccc@{}}
\toprule
& \textbf{BBC} & \textbf{Reddit} & \textbf{US cinemas}\\
\midrule
Platform type
  & Online newspaper & Social news & Cinema attendance\\
Period
  & Oct 2018 -- Jun 2019 & 2022 & 1983–2018\\
Items
  & 3{,}087 & $\sim$56{,}000 & 14,000\\
Mean activity/gross income & 276 & 41 & 17.7 million\\
$\hat\gamma_B - \hat\gamma_H$ & $\sim$2\% & $\sim$3\% & \\
Significant? & No ($p=0.67$) & No ($p=0.28$) & No \\
\bottomrule
\end{tabular}
\caption{Cross-platform comparison of the effect of hits. The effect of hits, as measured by the slowing factor $\gamma$, is small and non-significant in all three cases, consistent with elastic collective attention.}
\label{tab:crossplatform}
\end{table}

Results obtained in all three datasets are summarized in Table~\ref{tab:crossplatform}.

\section*{Methods}

\subsection{Empirical datatasets}
We regularly crawled the sport section of the BBC website (its front page and pages dedicated to individual sports) and collected news articles with comment sections. From October 1, 2018, to June 30, 2019, we collected 3,087 articles that received 852,400 comments from 67,527 readers. Each article is assigned to a sport category. The most populated categories are Football (1590 articles), Rugby Union (439 articles), Cricket (240 articles), Tennis (162 articles), Formula 1 (139 articles), Golf (123 articles) and Boxing (103 articles). Each comment is time-stamped at one-minute resolution. BBC typically closes comments at the second midnight after an article is published; most are therefore open for 24--48 hours. This dataset was extensively analyzed in~\cite{medo2022simple}.

We obtained comments on Reddit and submissions from the AcademicTorrents mirror \enquote{Reddit comments/submissions 2005-06 to 2025-06} \cite{academictorrents_reddit2005_2025}.  The underlying data comes from the Pushshift Reddit archival dataset \cite{baumgartner2020pushshift}, followed by a mirror with updates assembled, split, and seeded by community maintainers (e.g.\ u/Watchful1, u/RaiderBDev).  We processed the data as ndjson, decompressed using zstandard, and filtered to months through 2022.  We note known coverage gaps in the original archives \cite{gaffney2018caveat}. There are approximately 56,000 posts and more than 2,300,000 comments on them. The comment distribution is broad and can be described by a power law with exponent $\alpha \approx 1.6$ across a broad range of popularity (Figure~\ref{fig:reddit}F). While the median post receives only 1~comment (mean $41.1$), the top $10\%$ of posts attract more than $90\%$ of all comments. This extreme inequality defines Reddit's attention economy.

We examined the \textit{worldnews} community over the first quarter of 2018, which comprised approximately 56,000 posts and 2.3~million comments.  Given the extremely skewed distribution of engagement (the median post received only one comment, while the top $10\%$ of posts attracted more than $90\%$ of all comments), we defined hit posts as the 25 most commented submissions within the three-month period. For each hit post, we extracted a time window centered on its surge of activity, aligned all events so the peak of comment activity occurred at time zero, and measured both the hit's trajectory and the background activity generated by all other posts during the same window.  This alignment enabled us to construct aggregate profiles of comment dynamics around news events.  In parallel, we sampled baseline intervals without major peaks to provide a control for ordinary commenting dynamics.  Across both hit and non-hit intervals, we recorded comment density at minute-level resolution, allowing us to characterize the rise and subsequent decay of activity with precision.  The resulting data set included 50 trajectories in total (25 hits and 25 corresponding baselines), which formed the basis for quantitative modeling of engagement dynamics.

We obtained the movie box office data for years 1983--2018 obtained from \url{https://www.boxofficemojo.com/}. As cinema attendance typically strongly varies during the week, we suppressed this effect by using weekly box office data instead of the initially obtained daily data. Upon data cleaning, 14,000 movies are present in the dataset. As both total weekly gross as well as the weekly gross of the most successful movies grow over time, we select the hit movies based on their gross revenue's percentile within the year when the movie was released.

\subsection{Measuring the effect of hits on other articles}
Our approach evaluates how the number of new comments articles receive changes over time. In particular, we count the number of new comments received by article $i$ in the window of length $\Delta t$ just before time $t_j$, $\Delta c_i^B(t_j)$, and the number of new comments in the window of length $\Delta t$ just after time $t_j$, $\Delta c_i^A(t_j)$. In the case of measuring the effect of hits, time $t_j$ is the time of appearance of a hit article. To measure general aging effects, $t_j$ is a randomly chosen time point. To avoid possible confounding effects of other articles, we consider only the time points $t_j$ for which no other article appears in the measurement period $[t_j-\Delta t, t_j + \Delta t]$.

To estimate the proportionality factor $\gamma$, we assume that the numbers of new comments are drawn from the Poisson distribution with some unknown value of mean activity $\mu_i(t_j)$ in the time window before $t_j$, and mean activity $\gamma\mu_i(t_j)$ in the time window after $t_j$. The likelihood of the observed data $\mathcal{D}$ has the form
\begin{equation}
\mathcal{L}(\mathcal{D}\vert \boldsymbol{\mu}, \gamma) =
\prod_{i, t_j} P[\Delta c_i^B(t_j)\vert\mu_i(t_j)] \times P[\Delta c_i^A(t_j)\vert\gamma\mu_i(t_j)]
\end{equation}
where the product is over all measurement times $t_j$ and articles with open comment sections, $\boldsymbol{\mu}$ is the vector of all mean activity values, and $P(n\vert\mu)=\mu^n\mathrm{e}^{-\mu}/n!$ (the Poisson distribution).\footnote{When times $t_j$ are the appearance times of hit articles, the hit articles themselves are naturally excluded from the likelihood function as the goal is to measure their effect on the \emph{other} articles.}
This likelihood function can be maximized analytically, leading to the maximum likelihood estimate $\hat\gamma$ given by \eref{gamma_est}. The confidence intervals for $\hat\gamma$ can be estimated by non-parametric bootstrap.

Denoting the standard deviations of the bootstrap estimates as $\sigma_B$ and $\sigma_H$ for the baseline and hit scenario, respectively, the significance of the observed difference between $\hat\gamma_B$ and $\hat\gamma_H$ can be assessed by computing its $z$-score
\begin{equation}
z = \frac{\hat\gamma^B-\hat\gamma^H}{\sqrt{\sigma_B^2 + \sigma_H^2}}
\end{equation}
and the corresponding two-tailed $p$-value. See SI for derivation details, description of the bootstrap procedure, and a comparison of the MLE estimate with other ways to estimate the proportionality factor $\gamma$. Note that the proposed measurement of the effect of hits is related to the much-studied problem of measuring treatment effects in econometrics~\cite{imbens2015causal}.

\subsection{The FAE model with Fitness, Aging, and Elasticity}
We create a synthetic dataset over $T_S$ discrete time steps, with each step representing one minute. In each step, a new article appears with probability $p_n$. The fitness of article $i$, $\eta_i$, is drawn from a given fitness distribution. The expected number of new comments at time $t$ is
\begin{equation}
\label{elastic_demand}
C(t) = n_0 (1-\lambda) + \lambda X\sum_j \eta_j D_j(t-t_j)
\end{equation}
where $n_0$ is chosen so that the average number of comments reaches a desired value at $\lambda=0$. The elasticity of the collective attention is tuned by $\lambda\in[0, 1]$, where $\lambda=0$ and $\lambda=1$ correspond to a non-elastic case and a perfectly elastic case, respectively. The exponential aging factor $D_j(t-t_j)=\exp[-(t - t_j) / \Theta_j]$ is a function of the article aging timescale, $\Theta_j$, and the article appearance time, $t_j$. Note that the exponential aging can be explained by a simple model where each article is of interest to a fixed pool of readers (the pool's size is determined by the article's attractiveness to the readers) and every reader has a fixed probability of writing their comment (most readers comment in a discussion only once) per time unit~\cite{ishii2012hit}. Exponential aging can also be interpreted as a limiting case of the recent bi-exponential impact model~\cite{candia2019universal}. Finally, the multiplying factor $X$ is set so that $\overline{C(t)}$ is independent of the elasticity parameter $\lambda$, thus ensuring that varying $\lambda$ leaves the data volume approximately unchanged. The actual number of new comments at time $t$ is drawn from the Poisson distribution with mean $C(t)$.

The probability that a single comment at time $t$ is added to article $i$ has the usual form~\cite{kong2008experience,medo2011temporal,wang2013quantifying}
\begin{equation}
\label{competition}
P(i, t) = \frac{\eta_i D(t - \tau_i)}{\sum_j \eta_j D(t - \tau_j)}.
\end{equation} nodes
The choice $\lambda=0$ in \eref{elastic_demand}, which we refer to as the non-elastic case, induces the much-studied network dynamics where the articles ``compete'' for the incoming links. By contrast, the choice $\lambda=1$, the perfectly elastic case, results in Eqs.~(\ref{elastic_demand}) and (\ref{competition}) yielding the expected number of new comments of article $i$ in the form $\overline{\Delta c_i(t)} = X\eta_i D(t-t_i)$, which is independent of the other articles' fitness values and appearance times. Here, we focus on article-level commenting dynamics and avoid modeling the user side (\ie, which user authored an individual comment). This decision is further supported by the lack of structure found in users' commenting patterns (see Fig.~S9 in SI).

To match the BBC data as closely as possible, we choose the same duration, $T_S=393,120$ (with an additional initial 2,000 steps to equilibrate the simulation), $p_n=7.9\cdot10^{-3}$, aging timescale values $\Theta_i$ distributed uniformly in the range $[150, 600]$, and $X=0.85$. The values of $R_i:=\eta_i\Theta_i$ are drawn from a combination of two exponential distributions, $\rho(R)=\tfrac12\exp(-R/450)/450+\tfrac12\exp(-R/150)/150$, which represent the more and less popular article categories, respectively. For given $R_i$ and $\Theta_i$, article fitness is obtained as $\eta_i=R_i/\Theta_i$. The underlying user activity $n_0$ changes on a daily basis; we obtain it as $1+c_1+c_2$, where $c_{1,2}$ are distributed uniformly in the range $[0,1$], thus leading to $\overline{n_0}=2$, which is close to the average number of comments per minute ($2.2$) in the BBC data. These parameters were used to obtain panels D--F in Figure~\ref{fig:fig1}.

\section*{Discussion}
Many real datasets feature a broad distribution of item popularity, giving rise to hit items that far outweigh the mean popularity. Here, we take advantage of rapid popularity growth in some datasets, which lets us measure the impact of newly emerging hits on the rest of the system. In the short window after their appearance, these hits often attract more attention than all other available items combined. Based on the common theories of limited attention~\cite{qiu2017limited,lorenz2019accelerating,candia2019universal,lazer2020studying}, the high attention given to the hits should in turn lead to reduced attention elsewhere---the hits should impact the rest of the system. Contrary to expectations, we find a small or negligible impact of hits in all three analyzed datasets. Another novel finding is transient background suppression in the Reddit data: At the moment of peak hit activity, the comment rate on non-hit posts drops by ${\sim}40\%$, recovering within minutes.  This brief reallocation of attention suggests that competition for collective attention operates on very short timescales ($<$5~min) before the system re-equilibrates. The slowing factor $\hat\gamma$, which averages over $\Delta t = 10$~min windows, does not capture this transient effect, which may motivate extensions of the FAE model with a short-timescale competition term.

Our study has limitations. The slowing factor is particularly suitable for platforms where hits reach high activity quickly, making it possible to compare the activity devoted to the rest of the system shortly before and shortly after the hit appears. It is an open question how to generalize our framework to systems where popularity builds up gradually either due to preferential attachment~\cite{wang2013quantifying} or a gradual build-up of expectations before an event~\cite{desiderio2025highly}. Some systems are granular in the sense that they include items of different types---news on politics and football, for example. Because an important political news item can have limited impact outside its field, the measurement's sensitivity can be improved by focusing only on items of the same type. This has two disadvantages: One must choose or identify suitable item types and their granularity for the measurement, and statistical uncertainty can grow when only part of the system is considered for analysis. While we used three distinct datasets as a basis for our analysis, future work may examine whether similar elasticity of collective attention holds for different kinds of cultural products, economic goods, and services. A comparison with existing approaches~\cite{berry2021leadership} could be possible in some cases.

To conclude, our findings contribute to the literature on popularity dynamics~\cite{kong2008experience, hofman2017prediction,fortunato2018science,yucesoy2018success} and collective attention~\cite{wu2007novelty,weng2012competition,parolo2015attention,plata2021neutral,ojer2026modeling} by demonstrating that there is a limit to the generality of widely observed patterns and mechanisms (such as preferential attachment). While previous studies have emphasized the generality of observed patterns of popularity and impact~\cite{candia2019universal,medo2022simple}, future research might focus more on identifying violations of pervasive patterns and the causes behind them. Moreover, because managing and influencing the spread of online information is vital for online newspapers and social platforms, our models and methods can inform decisions by newspaper editors and content creators.

The finding that the decay of attention upon hit appearance is comparable to the normal aging in the absence of new competition suggests that the competition for collective attention is weaker than previously thought~\cite{gleeson2016effects,plata2021neutral}. This has important consequences for e-commerce, where limited attention can be an important decision criterion. In the movie industry, for example, the timing of movie releases with respect to competitors is an important consideration~\cite{krider1998competitive,natividad2013financial}. Measuring the short-term impact as well as the long-term compound impact of a strong competitor is particularly valuable in this context.

\section*{References}
\bibliographystyle{naturemag}
\bibliography{references_competition}

\end{document}